\documentclass[5p,sort&compress]{elsarticle}
\usepackage{amssymb,amsmath,amsfonts,psfrag}

\usepackage{dsfont}
\usepackage{graphicx}

\newcommand{\unitop}{\mathds{1}}
\newcommand{\bra}[1]{\left\langle #1 \right|}
\newcommand{\ket}[1]{\left| #1 \right\rangle}
\newcommand{\Tr}[1]{\mathrm{Tr} \left[ \, #1 \, \right]}

\newcommand{\bea}{\begin{eqnarray}}
\newcommand{\eea}{\end{eqnarray}}
\newcommand{\be}{\begin{equation}}
\newcommand{\ee}{\end{equation}}

\newcommand{\nn}{\nonumber}
\newcommand{\onehalf}{\frac{1}{2}}

\newcommand{\figref}[1]{Fig.~\ref{#1}}

\def\BA{\begin{eqnarray}}
\def\BE{\begin{equation}}
\def\EA{\end{eqnarray}}
\def\EE{\end{equation}}

\def\gtsim{\lower-0.45ex\hbox{$>$}\kern-0.77em\lower0.55ex\hbox{$\sim$}}
\def\ltsim{\lower-0.45ex\hbox{$<$}\kern-0.77em\lower0.55ex\hbox{$\sim$}}

\begin{document}

\title{A New Approach to the  Gluon Structure Function}

%-------------------------------------------------------------------
\author[Heidelberg]{D.~Gr\"unewald}
%\ead{daniel@tphys.uni-heidelberg.de}
%
\author[Berlin]{E.-M.~Ilgenfritz}
%\ead{ilgenfri@physik.hu-berlin.de}
%
\author[Heidelberg,Heidelberg2]{H.J.~Pirner}
%\ead{pir@tphys.uni-heidelberg.de}
%

\address[Heidelberg]{Universit\"at Heidelberg, Institut f\"ur Theoretische Physik, 69120 Heidelberg, Germany}
\address[Heidelberg2]{Max-Planck-Institut f\"ur Kernphysik, 69117 Heidelberg, Germany}
\address[Berlin]{Humboldt-Universit\"at zu Berlin, Institut f\"ur Physik,
  12489 Berlin, Germany}

\date{\today}

\begin{abstract}
%---------------
We calculate the gluon structure function of a color dipole in 
a new approach evaluating the matrix elements 
of $SU(2)$ gluon field operators separated along a direction close to the 
light cone. As vacuum state in the pure glue sector, we use 
a variational ground state of the near-light-cone Hamiltonian.
With a mean momentum fraction of the gluons fixed to the "experimental 
value" in a proton, the resulting gluon structure function for a dipole 
state with four links is compared qualitatively to the NLO \emph{MRST} 2002
parameterization at $Q^2=1.5\,\mathrm{GeV}^2$.
\end{abstract}


\maketitle

%% Begin Document

In Euclidean lattice simulations, 
using the operator product expansion 
the lowest moments of the meson and nucleon structure functions have been 
evaluated~\cite{Dolgov:2002zm,Gockeler:2004wp,Gockeler:1996zg,Meyer:2007tm}. 
On the other side,  
Loop-loop correlation functions of tilted Wegner-Wilson 
loops computed in Euclidean space~\cite{Shoshi:2002rd,Giordano:2008ua} 
can be related to the gluon distribution function of a color dipole 
which consists of a static quark and antiquark pair connected by a 
Schwinger string. Alternatively, an approach based on
light cone dynamics~\cite{Naus:1997zg,Dalley:2003aj} 
refers to constituents moving along the light cone, as suggested by the 
picture of Feynman scaling (as zeroth approximation). 
In the light cone approach the non-perturbative QCD vacuum structure 
is hard to achieve within the Fock representation of free fields acting 
on a trivial vacuum. 
As we will demonstrate, the confining nontrivial QCD vacuum is essential to 
generate the correct interaction of  colored constituents moving along
light like trajectories.    

Therefore 
we have developed a near-light-cone (nlc) approach in which we exploit
%% the $SU(2)$ lattice formulation, 
the lattice formulation given for $SU(2)$ gluodynamics, 
benefitting from simplifications emerging 
in the light cone limit.
In Ref.~\cite{Grunewald:2007cy} we have constructed a ground state wave 
functional of the nlc Hamiltonian which is simpler than the ground state 
in equal-time theory. 
In Ref.~\cite{Grunewald:2009zv} we have outlined the formalism to 
determine the gluon distribution function 
of a color dipole with this  
ground state wave functional. The present letter gives the main new results
which one obtains following this approach.

The gluon structure function $g(x_B)$ is the probability that a gluon carries 
a fraction $x_B$ of the longitudinal momentum of the fast moving hadronic target. 
In light cone coordinates, it is given by the Fourier transform of the matrix 
element of the two-point operator $G_{lc}(z^-,\vec{z}_\perp\,;0,\vec{z}_\perp)$ of 
longitudinally separated gluon field strength operators in a hadron state 
$|h(p_-,\vec{0}_\perp)\rangle$:
\begin{multline}
g(x_B)%&
=
%&
\frac{1}{x_B}\,\frac{1}{2\,\pi}\int_{-\infty}^{\infty} dz^-\,d^2\vec{z}_\perp%\,dz^\prime{}^{-}
\,e^{-\mathrm{i}\,x_B\,p_-\,z^-}\,\frac{1}{p_-} \\
%& &
\bra{h(p_-,\vec{0}_\perp)}G_{lc}(z^-,\vec{z}_\perp\,;0,\vec{z}_\perp)
\ket{h(p_-,{\vec 0}_\perp)}_c \, , 
\label{eq:eq1}
\end{multline}
with 
\begin{multline}
G_{lc}(z^-,\vec{z}_\perp\,;\,0,\vec{z}_\perp)=\sum\limits_{k=1}^2 F_{-k}^a(z^-,\vec{z}_\perp) \\
\times\,S_{a b}^A(z^-,\vec{z}_\perp\,;\,0,\vec{z}_\perp)\,F_{-k}^b(0,\vec{z}_\perp) \, .  
\label{eq:eq2}
\end{multline}
The hadron $|h(p_-,\vec{0}_\perp)\rangle$ 
is centered in transversal configuration space at 
$\vec{b}_\perp=\vec{0}_\perp$ and carries a longitudinal momentum $p_-$.
The index ``c'' indicates that the connected matrix element has to be taken.
The Schwinger line \newline $S_{ab}^A(z^-,\vec{z}_\perp\,;\,0,\vec{z}_\perp)$ in the adjoint representation and
running along a light like path is inserted between the gluon field strength
operators $F_{-k}^b(0,\vec{z}_\perp)$ and   $F_{-k}^a(z^-,\vec{z}_\perp)$.
The importance of the Schwinger lines along the light cone has been demonstrated
e.g. in the loop-loop correlation model where hadron-hadron scattering 
cross-sections are calculated from Wegner-Wilson loop correlation 
functions~\cite{Shoshi:2002rd}. In another language, the eikonal phases arising from the strings 
along the $x^-$- direction describe ``final state'' interaction 
effects which distinguish structure functions from parton 
probabilities~\cite{Brodsky:2002ue}.

We are using near-light-cone coordinates which allow us to
implement light front quantization as a limit of equal time quantization. 
The definition of the temporal nlc coordinate $x^+$ contains an additional 
external parameter $\eta$ which facilitates a smooth interpolation between 
equal time quantization ($\eta=1$ , $x^+= x^0$) and light cone quantization 
($\eta=0$ , $x^+=1/2\,( x^0+x^3)$). 
The definitions are: 
\begin{equation}
x^+=\frac{1}{2}\Bigl[\left(1+\eta^2\right)x^{0}+ \left(1-\eta^2\right)x^{3}\Bigr] ~,~
x^-=\phantom{\frac{1}{2}}\Bigl[x^{0}-x^{3}\Bigr] \; . \nonumber
\end{equation}
The $\eta \to 0$ limit can be interpreted as the infinite momentum frame 
limit in which the partons of the color dipole move with infinite momentum. 
The use of nlc coordinates has the advantage that no quantum constraint 
equations have to be solved. 

In the $SU(2)$ lattice formulation we have two transversal gauge fields $A_j$ ($j=1,2$) 
and one longitudinal one $A_-$ ($j=-$) that are represented by the corresponding 
link matrices $U_j(\vec{x})$. The $A_+$ component of the gauge field is set equal 
to zero. As a result one has the Gauss-law constraining the entire Hilbert space
to the physical sector of gauge invariant states. The gluon dynamics is determined 
by the Hamiltonian which has been derived in Ref.~\cite{Grunewald:2007cy}. 
It represents the gluon energy density on the lattice. The QCD coupling constant 
enters as $\lambda=4/g^4$ in the $SU(2)$ case,
\begin{multline}
\mathcal{H}_{\mathrm{eff,lat}}= \frac{1}{N_-N_{\bot}^2}\frac{1}
{a_{\bot}^4} \frac{2}{\sqrt{\lambda}}
\sum\limits_{\vec{x}}\left\{
    \frac{1}{2}~\sum\limits_{a}~\Pi_-^a(\vec{x})^2~
    +
    \right.  \\  
    \onehalf~\lambda~\Tr{\unitop-U_{12}(\vec{x})}
    ~+\sum\limits_{k,a}~\frac{1}{2}~\frac{1}{\tilde{\eta}^2}~ 
      \Bigg[~\Pi_k^a(\vec{x})^2+\Bigg.\\
     \left.\Bigg.
      \lambda~\Biggl(\Tr{\frac{\sigma_a}{4i}
      ~\left(U_{-k}(\vec{x})-U_{-k}^{\dagger}(\vec{x})\right)}\Biggr)^2
      \Bigg] \right\}\; .
\label{eq:eq4}
\end{multline}
Here, $U_{12}$ are purely transversal plaquettes, and $U_{-k}$ are 
longitudinal-transversal plaquettes.  

The Hamiltonian contains the chromo-electric field strength operators 
$\Pi_i^a(\vec{x})$, which are canonically conjugate to the links $U_i(\vec{x})$.
They obey commutation relations which follow from the corresponding continuum 
relations,
\be
\left[\Pi_i^a(\vec{x}),U_j(\vec{y}) \right]= 
\left(\sigma^a/2\right)\,U_i(\vec{x})\,\delta_{\vec{x},\vec{y}}\,\delta_{i,j}\,.
\label{eq:eq5}
\ee
The constant $\tilde{\eta}$ in Eq.~(\ref{eq:eq4})    
is the product $\tilde{\eta} = \eta \cdot \xi$ of the near-light-cone parameter 
$\eta$ and an eventual anisotropy parameter $\xi=a_-/a_\bot$, the ratio of 
lattice spacings in longitudinal and transverse directions.
If one chooses $\eta=1$ and varies $\xi$, one simulates an anisotropic equal 
time theory. In the limit $\xi \to 0$ one ends up with a system, which is 
contracted in the longitudinal direction. Verlinde and 
Verlinde \cite{Verlinde:1993te} and Arefeva \cite{Arefeva:1993hi} 
have advocated such a lattice to describe high energy scattering. 
A longitudinally contracted system means that even the minimal momenta become 
high in longitudinal direction. The limit $\xi \to 0$ leads to the same physics 
as the light cone limit $\eta \to 0$ with isotropic lattice spacings in 
longitudinal and transverse directions.
In both cases the nlc Hamiltonian is dominated by the terms proportional 
to $1/\tilde{\eta}^2$ involving transverse chromo-electric and chromo-magnetic 
fields. For a nlc-Hamiltonian formulation, the two-point operator in Eq.~(\ref{eq:eq2}) has to be
replaced by
\begin{multline}
G_{lc}(z^-,\vec{z}_\perp\,;\,0,\vec{z}_\perp)=\frac{1}{2}\, \sum\limits_{k=1}^2 
\left( F_{-k}^a(z^-,\vec{z}_\perp) \right. \\
\left. \times\,S_{a b}^A(z^-,\vec{z}_\perp\,;\,0,\vec{z}_\perp)\,\Pi_k^b(0,\vec{z}_\perp) + 
\left\{ F_{-k}^a \leftrightarrow \Pi_k^a \right\} \right) \, .  
\label{eq:eq6}
\end{multline}

In Ref.~\cite{Grunewald:2007cy} we have determined a variational gluonic 
ground state wave functional $\ket{\Psi_0}$ which consists of a product of 
single-plaquette wave functionals with 
two variationally optimized parameters $\rho_0=\rho_0(\lambda,\tilde{\eta})$ and $\delta_0=\delta_0(\lambda,\tilde{\eta})$,
\be
\ket{\Psi_0}=\Psi_0[U]\,\ket{0}=\sqrt{N_{\Psi}}\,e^{f[U]}\,\ket{0} \, , \nn
\label{eq:eq7}
\ee
\begin{equation}
f[U]=\sum\limits_{\vec{x}}
               \left\{\sum\limits_{k=1}^2 
                 \rho_0\,\Tr{U_{-k}(\vec{x})}
                +\delta_0\,\Tr{U_{12}(\vec{x})}
               \right\} \; .\label{eq:eq8} %\nn\\
           %& & \phantom{1}    
\nonumber
\end{equation}
$N_{\Psi}$ is a normalization factor. Here, the state $\ket{0}$ represents 
the trivial ground state which is annihilated by the field momenta 
$\Pi_k^a(\vec{x})$ conjugate to the links,
\be
\Pi_k^a(\vec{x})\,\ket{0}=0\,\,\mathrm{and}\,\,\bra{0}\,\Pi_k^a(\vec{x})=0\,\,~\mathrm{for}~\forall\,\vec{x},k,a\,.
\label{eq:eq9}
\ee
We have optimized this ansatz and extrapolated the parameters $\rho_0,\delta_0$ 
to the light cone $\tilde{\eta} \to 0$ (see Ref. \cite{Grunewald:2007cy}). 

\begin{figure}
\centering
\includegraphics[width=.5\textwidth]{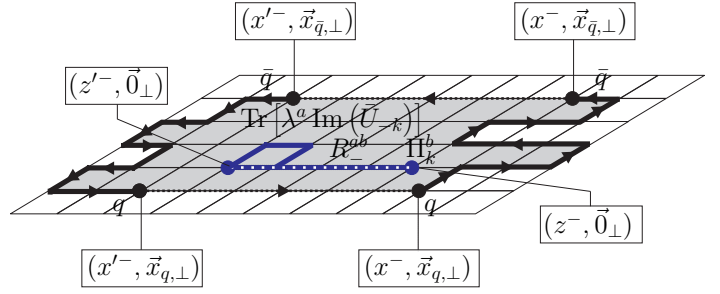}
\caption{The generalized Wegner-Wilson loop generating the matrix element
between the color dipole states. 
Quark $q$ and antiquark $\bar q$ are connected by $n$ transversal links. 
Only one transversal direction is shown.
The insertion represents the gluonic two-point operator with an electric 
and a magnetic field operator.}
\label{fig:Figure1}
\end{figure}

Since the nlc Hamiltonian in Eq.~(\ref{eq:eq4}) contains only gluon fields, 
we cannot {\it derive} a full hadronic wave function from this 
Hamiltonian. We have to make a model taking care of the gluon structure alone
while treating the quarks schematically. Our model represents a dipole 
localized in transversal configuration space at a fixed center of mass 
position $\vec{b}_\perp=\vec{0}$.
The distance $\vec{d}_\perp$ between quark 
and antiquark is bridged by a Schwinger line along some path $\mathcal{C}_\perp$ 
in the transversal plane. The in and out dipole states form a Wegner-Wilson 
loop in the eikonal approximation as sketched in \figref{fig:Figure1}.  

The longitudinal lattice momenta must be $n$-fold (integer) multiples 
of $2\,\pi/N_-$, with $0 \leq n\leq N_-/2-1$, since the longitudinal light 
cone momentum for an on shell particle is always positive.
The momentum $p_-$ of the target is chosen (in lattice units) as the largest 
momentum in order to allow for the maximum resolution in the gluon distribution
function \cite{Burkardt:2001jg}.
The longitudinal lattice {\it gluon momenta} then have the resolution
\be
p_-=\frac{2\,\pi}{N_-}(\frac{N_-}{2}-1)~~,~~\frac{\Delta p_-^g}{p_-}=\frac{2}{N_--2}\,.
\ee
In order to have a high resolution, the extension of the lattice in the 
longitudinal direction has to be very large. 

We impose the quark dynamics of the color dipole
externally. Since the total hadron longitudinal momentum is given by
the sum of the momenta of its constituents (quarks$+$gluons), 
the typical mean gluon momentum is taken from experiment. 
For a rough qualitative comparison, we use the \emph{MRST}-parameterization \cite{Martin:2002aw} of
the proton $SU(3)$ gluon distribution function
at the input scale $Q^2\approx\pi^2/a_\perp^2=1.5\,\mathrm{GeV}^2$ corresponding to $\lambda\approx10$, 
and assign a mean momentum fraction $p_-^S=0.38\,p_-$ to the gluon system of the $SU(2)$ color dipole. 
 
The gluon distribution function for a one-link dipole $ p_-^g\,g_1(p_-^g;p_-^S)$ 
with total gluonic momentum $p_-^S$ has been computed on lattices with 
$N_-=20$, $30$, $50$, and $100$ sites in the longitudinal direction.  
\begin{figure}[h]
\centering
\includegraphics[width=.5\textwidth]{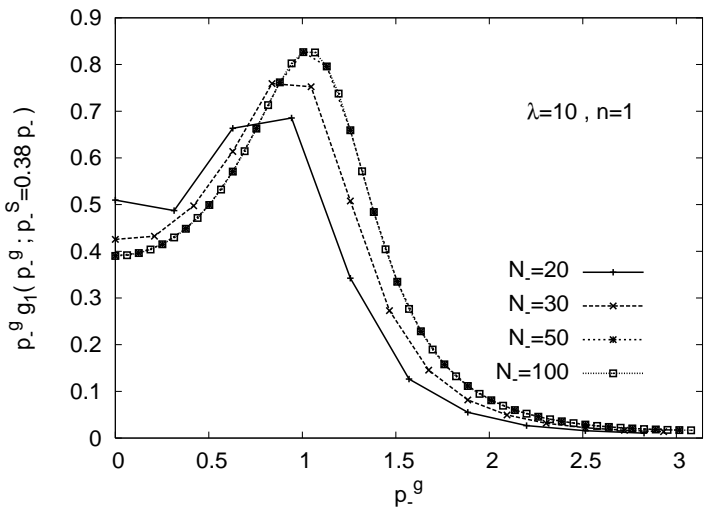}
\caption{Gluon distribution function $p_-^g\,g_1(p_-^g\,;\,p_-^S=0.38\,p_-)$ 
for a single transversal link, $n=1$.
The dependence on the longitudinal lattice size $N_-$ is shown for
a lattice coupling $\lambda=4/g^4=10$ (see Eq.(4)). 
The average gluon momentum $\langle p_-^g\rangle=p_-^S$ was input and
adjusted to $\langle x_B \rangle~p_-=0.38~p_-$ (see text).}
\label{fig:Figure2}
\end{figure} 
\figref{fig:Figure2} demonstrates the effect of increasing the number of 
longitudinal lattice sites, i.e. approaching the infinite volume
limit. Scaling for $\lambda=10$ seems to be obeyed for longitudinal lattice
extensions larger than $N_-=50$. Realistic lattice simulations need quite large 
longitudinal lattice sizes. 
The smearing of the distribution function is due to the gluon dynamics 
incorporated in the Wegner-Wilson loop expectation value. Thus, 
the area law behavior of the Wegner-Wilson loop yields a non-trivial 
gluon wave function which broadens the distribution.

For larger $\lambda$, i.e. smaller QCD gauge coupling $g^2$, 
the peak in the one-link distribution function becomes narrower.
In the extreme weak coupling limit, when the link reduces to a single gluon, 
the gluon distribution function becomes totally sharp.
On the other hand, for strong coupling one has a broad momentum distribution 
peaked around $p_-^S$.

The one-link dipole gluon distribution is the basic building block from which 
the multiple link dipole gluon distribution function of a hadron can be constructed.  
The actual hadronic state arises from a superposition of multiple link 
configurations. Let the wiggly strings $S_{q\,\bar{q}}$ (c.f. \figref{fig:Figure1})
connecting the quark and antiquark have a fixed number of transversal links 
since all the dipole configurations with fixed transversal length $n$ have the 
same energy. Then one
rotates the hadron in the transversal plane by summing over randomly chosen 
curves $\mathcal{C}_\perp$,  
in order to project this state on angular momentum $J_z=0$. 
>>From the random walk follows that for $n$-links , 
the hadron has an average radius squared $\vec{R}_\perp^2$ proportional to $n$: 
Hence, the area of the hadron scales with the number of links:
\be
\left\langle R_\perp^2 \right\rangle = n\,a_\perp^2/2~\, .
\label{eq:eq11}
\ee
Due to small $\delta_0$ in the light cone limit, the ground state wave functional
allows for ordinary strong coupling methods. This implies especially that one needs
incoming and outgoing states sharing the same transversal links connecting the 
quark and antiquark.
Therefore, the gluon  distribution can be obtained for a special string elongated 
along only one of the transversal axes (c.f. \figref{fig:Figure1}).

The computation of the n-link gluon distribution function 
$g_n$ is done in analogy to the computation of the single 
link gluon distribution. In strong coupling the total loop factorizes,
therefore  the $n$-link distribution function is given by 
the product of a splitting function $P_{n\rightarrow n-1}$ multiplying the gluon
distribution function with $n-1$ links. In the emerging recursion relation 
all possible intermediate momenta of the substring are summed over: 
\be
g_n\big(p_-^g;p_-^S \big)=
\frac{2\,\pi}{N_-}\sum_{p_-^S{}^\prime=0}^{p_-^S}g_{n-1}\big(p_-^g;p_-^S{}^\prime\big)\,P_{n\rightarrow
n-1}(p_-^S,p_-^S{}^\prime)
\label{eq:eq12}
\ee
The splitting function  $P_{n\rightarrow n-1}(p_-^S,p_-^S{}^\prime)$
denotes  the probability that a string with $n$ transversal links and total 
momentum $p_-^S$ splits into a string with $n-1$ transversal links
and total momentum $p_-^S{}^\prime$ and is given in Ref. \cite{Grunewald:2009zv}. 
The initial condition for the recursion relation Eq.~(\ref{eq:eq12}) 
is given by the one-link dipole function $g_1(p_-^g;p_-^S)$. 
The computation is purely arithmetic for small $\delta_0$,
and we can use a large longitudinal lattice with $N_-=1000$ lattice sites.
If one increases the number of transversal links, the gluons have
access to a larger region in phase space due to the splitting function 
$P_{n\rightarrow n-1}$ in Eq.~(\ref{eq:eq12}). Therefore the total gluon momentum 
will be partitioned among more gluons.  Hence, it becomes more likely to find
a gluon with a small fraction of the total momentum. 

\begin{figure}
\centering
\includegraphics[width=.5\textwidth]{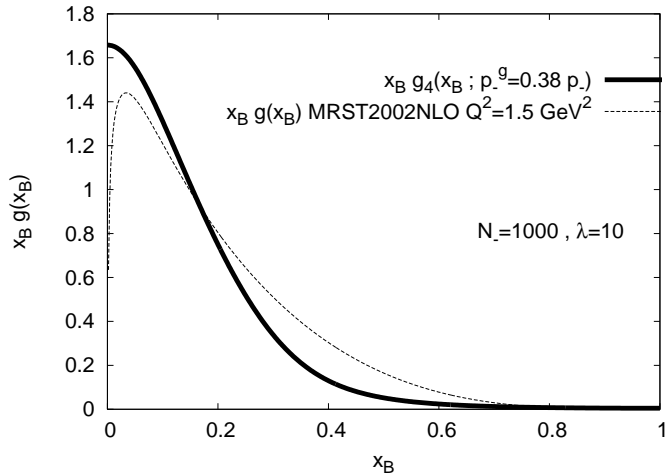}
\caption{Qualitative comparison of the gluon distribution function $x_B g_n(x_B;p_-^g)$ 
for $n=4$ transversal $SU(2)$ links with the \emph{MRST} gluon distribution function of
a proton at $Q^2 = 1.5 {\rm~GeV}^2$. 
The average gluon momentum was taken as input for the lattice
calculation on a $N_-=1000$ lattice at gauge coupling $\lambda=10$.}
\label{fig:Figure3}
\end{figure}
In \figref{fig:Figure3}, we show the theoretical
gluon structure function for a
$n=4$ $SU(2)$ link dipole as a function of the gluon fractional momentum $x_B=p_-^g/p_-$. 
To have a rough qualitative comparison, we also show 
the \emph{MRST} gluon distribution function at $Q^2=1.5\,\mathrm{GeV}^2$ which is
for real protons and $SU(3)$. 
As before, the first 
moment of the lattice gluon distribution function has been fixed 
in this figure to the value $\langle x_B\rangle=0.38$ at 
$Q^2=1.5\,\mathrm{GeV}^2$. We choose four links to be 
consistent with the size of a proton and the relation 
$\langle R_\perp^2 \rangle = n\,a_\perp^2/2~$
and a transversal lattice spacing of $a_\perp \approx 0.65~\mathrm{fm}$.
The lattice gluon distribution function for a color dipole 
shows a similar behavior as the phenomenological \emph{MRST}-gluon distribution 
function for a proton. It has been shown
in the loop-loop correlation model \cite{Shoshi:2002in}
that it is a quite good approximation to consider the three valence quarks
in the proton to be arranged as a dipole with a diquark and a quark.

The model presented here also shows that $x_B\, g(x_B) $ for the gluon at small 
$x_B$ becomes proportional to the hadronic size $ R_\perp^2$. This coincides 
with the empirical soft Pomeron behavior of hadronic cross sections. To evolve 
the structure function with increasing resolution 
$Q^2$ and with decreasing $x_B$ in Hamiltonian lattice QCD, 
one needs a more sophisticated ground state wave 
functional 
respecting scaling with the lattice spacing.

%% End Document

%------------------------------------------------------------------------

\end{document}